\documentclass[aps,preprint,groupedaddress,showpacs]{revtex4}
\usepackage{amsmath}
\usepackage{amssymb}
\usepackage{amsfonts}
\usepackage[dvips]{graphicx}
\usepackage[]{caption}
\begin{document}

\title{Effective interactions between star polymers\\ and colloidal particles}

\author{A. Jusufi}
\email[e-mail address: ] {jusufi@thphy.uni-duesseldorf.de}
\affiliation{Institut f{\"u}r Theoretische Physik II,
Heinrich-Heine-Universit\"at D{\"u}sseldorf,
Universit\"atsstra{\ss}e 1, D-40225 D\"usseldorf, Germany}
\author{J. Dzubiella}
\affiliation{Institut f{\"u}r Theoretische Physik II,
Heinrich-Heine-Universit\"at D{\"u}sseldorf,
Universit\"atsstra{\ss}e 1, D-40225 D\"usseldorf, Germany}
\author{C. N. Likos}
\affiliation{Institut f{\"u}r Theoretische Physik II,
Heinrich-Heine-Universit\"at D{\"u}sseldorf,
Universit\"atsstra{\ss}e 1, D-40225 D\"usseldorf, Germany}
\author{C. von Ferber}
\affiliation{Institut f{\"u}r Theoretische Physik II,
Heinrich-Heine-Universit\"at D{\"u}sseldorf,
Universit\"atsstra{\ss}e 1, D-40225 D\"usseldorf, Germany}
\author{H. L{\"o}wen}
\affiliation{Institut f{\"u}r Theoretische Physik II,
Heinrich-Heine-Universit\"at D{\"u}sseldorf,
Universit\"atsstra{\ss}e 1, D-40225 D\"usseldorf, Germany}
\date{\today}

\begin{abstract}
Using monomer-resolved Molecular Dynamics simulations and theoretical
arguments based on the radial dependence of the osmotic pressure in
the interior of a star,
we systematically investigate the effective interactions between
hard, colloidal particles and 
star polymers in a good solvent. The relevant parameters are the
size ratio $q$ between the stars and the colloids, as well as the 
number of polymeric arms $f$ (functionality) attached to the common
center of the star. By covering a wide range of $q$'s
ranging from zero (star against a flat wall) up to about 0.75,
we establish analytical forms for the star-colloid interaction which
are in excellent agreement with simulation results. A modified 
expression for the star-star interaction for low functionalities,
$ f \lesssim 10$ is also introduced.  
\end{abstract}

\pacs{82.70.Dd, 61.20.Ja, 61.41.+e}

\maketitle

\section{Introduction}

The study of the structural and thermodynamic properties of the 
polymeric state of matter has a long history in physics, which started
with the pioneering work of Flory \cite{flory:42,flory:49,flory}.
In the traditional, ``polymeric approaches'' to the matter, the 
chain nature of the macromolecules involved is in the foreground.
However, in the last few years, alternative, complementary considerations
have emerged that can loosely be called ``colloidal approaches''.
Here, one envisions the polymer chains as diffuse, spherical objects
and the chain nature of the molecules does not explicitly
appear in the formalism. Instead, in a first step, almost all of the 
monomeric degrees of freedom are
thermodynamically traced out of the problem \cite{likos:physrep:01}.
Thereby, the polymers are
replaced either by their centers of mass or by one of the monomers
along their backbone, typically the end- or the central monomer.
In this way, effective interactions between the polymers 
naturally arise \cite{likos:physrep:01}, which implicitly include
the effects of the traced-out monomers and typically have the range
of the chain radius of gyration $R_{\rm g}$. The earliest such approach
dates back to the work of Asakura and 
Oosawa \cite{asakura:oosawa:54,asakura:oosawa:58}, and
Vrij \cite{vrij:76}, who modeled polymer chains as penetrable spheres.
These models pertain mostly to Gaussian, i.e., ideal chains and
are semi-quantitative.
More systematic approaches have appeared in the
recent years, in which self-avoiding chains are modeled and effective 
interactions among them are derived by means of 
simulations \cite{ard:peter:00} or theory \cite{fuchs:schweizer:epl:00}.
The gain from adopting such an alternative view is twofold: on the
one hand, one has the possiblity of looking at the same problem from
a different angle; on the other hand, tracing out the monomers 
reduces the complexity of the problem by a factor $N$, the degree of
polymerization of the chains \cite{bolhuis:jcp:00}.

A physical system where the colloidal approach finds an intuitive and
natural application is that of star polymers \cite{Grest:review:96}.
These macromolecular entities are 
synthesized by covalently attaching $f$ polymeric chains
on a common center. In this way, hybrid particles between 
polymers and colloids can be constructed, which naturally bridge the
gap between these two common states of soft matter. The number of arms
$f$, also known as {\it functionality} of the stars, allows us to 
go from free chains ($f = 1, 2$) to stiff, spherical particles
($f \gg 1$).
Effective interactions
between star polymers in good \cite{likos:etal:prl:98} and 
$\Theta$-solvents \cite{likos:theta:pre:98} have been recently
derived and the validity of the former has been confirmed through
extensive comparisons with 
experiments \cite{likos:etal:prl:98,stellbrink:ecis98,stellbrink:ecis00}
and simulations \cite{jusufi:macromolecules:99,shida:00}. Extensions
to polydisperse stars \cite{ferber:polydisperse}
as well as to many-body forces in dense star polymer 
solutions \cite{Ferber:Jusufi:99:1} have also been recently 
carried out.

In this work, we wish to carry these considerations one step
further by looking at a {\it two-component} system of star polymers
in good solvent conditions and hard, spherical, colloidal particles.
Though the star-star interaction is readily available and the
colloids can be modeled as hard spheres, the effective cross interaction
between star polymers and colloids is still missing. It is the 
purpose of this work to present theoretical and simulation results and to
furnish analytic expressions for the force and/or the 
effective interaction acting between a star polymer and a spherical,
colloidal particle for a large range of size ratios between the two.
The theoretical approach is inspired by the eariler considerations
of Pincus \cite{pincus:porc:91} regarding the force acting between 
a star and a flat wall but are made more precise here and they are
also extended to include the effects of curvature. The rest of this
paper is organized as follows: in section \ref{Theory} we present
the general theoretical approach, both for flat and curved surfaces
and derive analytic expressions for the star-colloid force which
include a handful of undetermined parameters. In section
\ref{simulation} we compare those with the results of monomer-resolved
Molecular Dynamics simulations and determine the free parameters 
in order to achieve agreement between theory and simulation results.
In section \ref{revised} we present a modified version of the
star-star potential which is valid for very low arm numbers,
$f \lesssim 10$, and in section \ref{summary} we summarize and
conclude.

\section{Theory}
\label{Theory}
Let us first define the system under consideration and its
relevant parameters. We consider a collection of star polymers
with functionality $f$ and
hard, spherical colloidal particles, the interaction between the latter
species 
being modeled through the hard sphere (HS) potential. By considering
two isolated members of each species, i.e., one star and one colloid,
our goal is to derive the effective interaction between the two.
The colloids have a radius $R_{\rm c}$, which is 
a well-defined length scale. 

The stars, on the other hand, are 
soft, hairy balls without a sharply defined boundary and this leads to
some freedom in defining length scales characterizing their spatial extent.
The experimentally
measurable length scale that naturally arises from small-angle neutron-
or X-ray-scattering experiments (SANS or SAXS) is the
radius of gyration $R_{\rm g}$ of the stars and the associated
diameter of gyration $\sigma_{\rm g} = 2R_{\rm g}$. For the 
theoretical investigations on the subject, however, another length
scale turns out to be more convenient, namely the so-called
corona radius $R_{\rm s}$ of the star or the associated 
corona diameter $\sigma_{\rm s} = 2R_{\rm s}$. The corona radius
arises naturally in the blob model for the conformation of isolated
stars, introduced by Daoud and Cotton \cite{Daoud:Cotton:82:1}. 
According to the Daoud-Cotton picture, the bulk of the interior of
a star in good solvent conditions (and for sufficiently long arm chains),
consists of a region in which the monomer density profile $c(s)$
follows a power-law as a function of the distance $s$ from the star center,
namely:
\begin{equation}
c(s) \sim a^{-3}\left({{s}\over{a}}\right)^{-4/3}\bar v^{-1/3}f^{2/3},
\label{daoud1}
\end{equation}
with the monomer length $a$,
the excluded volume parameter $v$ and the reduced excluded volume parameter
$\bar v \equiv v/a^3$. Outside this scaling region, there exists a diffuse
layer of almost freely fluctuating rest chains, in which the scaling
behavior of the monomer profile is not any more valid. We define the
corona radius $R_{\rm s}$ of the star as the distance from the center
up to which the scaling behavior of the monomer density given by 
Eq.\ (\ref{daoud1}) above holds true. In what follows, we define the
{\it size ratio} $q$ between the stars and the colloids 
as:
\begin{equation}
q\equiv\frac{R_{\rm g}}{R_{\rm c}}.
\label{size.ratio}
\end{equation}
 
In addition, the interior of the star forms a semidilute polymer solution
in which scaling theory \cite{degennes} predicts that the osmotic pressure
$\Pi$ scales with the concentration $c$ as $\Pi(c) \sim c^{9/4}$. 
Combining the latter with Eq.\ (\ref{daoud1}) above, we obtain for the
radial dependence of the osmotic pressure of the star within the 
scaling regime the relation:
\begin{equation}
\Pi(s) \sim k_BT\,f^{3/2}s^{-3}\qquad\qquad(s\leq R_{\rm s}).
\label{daoud2}
\end{equation}
No relation for the osmotic pressure $\Pi(s)$ for the diffuse region
$s > R_{\rm s}$ is known to date. It is indeed one of the central points
of this work to introduce an accurate ansatz for the latter, one that
will allow us also to derive closed formulas for the effective force
between a star and a hard object. This is the subject we examine below.

\subsection{A star polymer and a flat wall}

We begin by examining the simplest case, in which a star center is
brought within a distance $z$ from a hard, flat wall, as depicted in
Fig. \ref{star-wall}. Going back an idea put forward some ten years
ago by Pincus \cite{pincus:porc:91}, we can calculate the force $F_{\rm sw}(z)$
acting between the polymer and the wall by integrating the normal
component of the osmotic pressure $\Pi(s)$ along the area of contact
between the star and the wall. In the geometry shown in Fig.\ \ref{star-wall},
this takes the form:
\begin{equation}
F_{\rm sw}(z) = 2\pi\int_{y = 0}^{y = \infty} \Pi(s)\cos{\vartheta}\,y{\rm d}y.
\label{fsw.1}
\end{equation}
Using $z = s\,\cos{\vartheta}$ and $y = z\,\tan{\vartheta}$ we can transform
Eq.\ (\ref{fsw.1}) into:
\begin{equation}
F_{\rm sw}(z) = 2\pi z\int_{z}^{\infty} \Pi(s){\rm d}s.
\label{fsw.2}
\end{equation}
Eq.\ (\ref{fsw.2}) above implies immediately that, if the functional
form for the force 
$F_{\rm sw}(z)$ 
were to be known, then the corresponding
functional form for the osmotic pressure $\Pi(z)$ could be obtained
through:
\begin{equation}
\Pi(z) \propto -\frac{{\rm d}}{{\rm d} z}
               \left(\frac{F_{\rm sw}(z)}{z}\right).
\label{piofz}
\end{equation}

To this end, we now refer to known, exact results regarding the force
acting between a flat wall and a {\it single, ideal chain} whose
one end is held at a distance $z$ from a flat wall 
\cite{eisenriegler:book}. There, it has been
established that the force $F_{\rm sw}^{\rm (id)}(z)$ is given by the relation:
\begin{equation}
F_{\rm sw}^{\rm (id)}(z) = k_BT\frac{\partial}{\partial z}
                \ln\left[{\rm erf}\left(\frac{z}{L}\right)\right],
\label{erich.1}
\end{equation}
where ${\rm erf}(x) = 2/{\sqrt{\pi}}\int_0^{x}e^{-t^2}{\rm d}t$ 
denotes the error function and
$L$ is some length scale of the order of the radius of gyration
of the polymer. Carrying out the derivative and setting $\rm erf(x) \cong 1$
for $x \gg 1$, we obtain a Gaussian form for the chain-wall force at
large separations:
\begin{equation}
F_{\rm sw}^{\rm (id)}(z) \cong \frac{k_BT}{L}
\exp\left(-\frac{z^2}{L^2}\right)\qquad\qquad(z \gg L).
\label{erich.2}
\end{equation}
We now imagine a star composed of ideal chains. As the latter do not interact
with each other (``ghost chains'') the result of Eq.\ (\ref{erich.2}) holds
for the star as well. Going now to self-avoiding chains, we assert that,
as the main effect giving rise to the star-wall force is the volume which
the wall excludes to the chains, rather than the excluded volume interactions
between the chains themselves, a relation of the form (\ref{erich.2}) 
must also hold for the force
$F_{\rm sw}(z)$ between a wall and a {\it real} star, 
but with the length scale $L$ 
replaced by the radius of gyration or the corona radius of the latter
and with an additional, $f$-dependent prefactor for taking into account
the stretching effects of the $f$ grafted polymeric chains.
From Eqs.\ (\ref{piofz}) and ({\ref{erich.2}) it now follows that
\begin{equation}
\Pi(s) \propto \frac{k_BT}{L}\left(\frac{1}{s^2} + \frac{2}{L^2}\right)
\exp\left(-\frac{s^2}{L^2}\right)\qquad\qquad(s \gg L).
\label{pressure.out}
\end{equation}

The full expression for $\Pi(s)$ now follows by combining Eq.\ (\ref{daoud2}),
valid for $s \leq R_{\rm s}$, with Eq.\ (\ref{pressure.out}), 
valid for $s \gg L \cong R_{\rm s}$, and matching them at $s = R_{\rm s}$.
The local osmotic pressure $\Pi(s)$ is the interior of a star polymer,
as a function of the distance $s$ from its center has hence
the functional form:
\begin{equation}
\label{osmotic.eq}
\Pi(s) = \Lambda f^{3/2}k_BT
\begin{cases}
           s^{-3} & \text{for $s \leq R_{\rm s}$};\\
                  \left(\frac{1}{s^2} + 2\kappa^2\right)
                        \frac{\xi}{R_{\rm s}}
                        \exp\left[-\kappa^2\left(s^2 - R_{\rm s}^2\right)\right]
                  & \text{for $s > R_{\rm s}$},
\end{cases} 
\end{equation}
where $\Lambda$ and $\kappa = L^{-1}$ are free parameters; it is to be
expected that $\kappa = O(R_{\rm g}^{-1})$, as we will verify shortly.
On the other hand, 
$\xi$ must be
chosen to guarantee that $\Pi(s)$ is continuous at $s = R_{\rm s}$, 
resulting into the value:
\begin{equation}
\label{xi}
\xi = \frac{1}{1 + 2\kappa^2 R_{\rm s}^2}.
\end{equation}

Eq.\ (\ref{osmotic.eq}) above concerns the radial distribution of the
osmotic pressure of an isolated star. The question therefore arises,
whether this functional form for the osmotic pressure can be used in
order to calculate the force between a star and a flat wall also in
situations where the star-wall separation is smaller than the radius
of gyration of the star, in which case it is intuitively expected that
the presence of the wall will seriously disturb the monomer distribution
around the center and hence also the osmotic pressure. In fact,
it is to be expected the osmotic pressure is a function of
{\it both} the star-wall separation $z$ and the radial distance
$s$, whereas in what follows we are going to be using Eq.\ (\ref{fsw.2})
together with Eq.\ (\ref{osmotic.eq}), in which $\Pi(s)$ has no
$z$-dependence itself. However, it turns out that this is an excellent
approximation. On the one hand, it is physically plausible for large
star-wall separations, where the presence of the wall has little
effect on the segment density profile around the star center and the
ensuing osmotic pressure profile. On the other hand, also
at very small 
star-wall separations, the scaling form $\Pi(s) \sim s^{-3}$ continues to
be valid. To corroborate this claim, we proceed with some arguments
to this effect.

First, we refer once more to known, exact results concerning the 
radial distribution of the pressure on a hard wall arising from an
ideal chain grafted on it \cite{bickel:etal:pre:00}, a situation
similar to holding one end of a chain at a distance very close to the 
wall surface. The pressure $\Pi_{\rm id}(s)$ reads as \cite{bickel:etal:pre:00}:
\begin{equation}
\Pi_{\rm id}(s) = \frac{1}{2\pi}\frac{1}{(s^2 + a^2)^{3/2}}
\left(1 + \frac{s^2 + a^2}{2R_{\rm g}^2}\right)
\exp\left[-\frac{s^2 + a^2}{4R_{\rm g}^2}\right],
\label{bickel}
\end{equation}
with the segment length $a$, indicating that in the regime
$a \ll s \ll R_{\rm g}$ indeed the scaling $\Pi_{\rm id}(s) \sim s^{-3}$
holds. 

Second, we can employ a scaling argument, asserting that, on dimensional
grounds, the osmotic pressure exerted by a star on a nearby 
flat wall and held at a distance $z$ from it, must be of the form 
$\Pi(s,z) = k_BT R_{\rm g}^{-3}\,h(s/R_{\rm g},z/R_{\rm g})$, 
with some scaling function
$h(x,y)$; universality arguments dictate that the segment length $a$
should not appear in the dimensional analysis and hence $s$, $z$ and
$R_{\rm g}$ are the only relevant length scales for this problem. 
Now, for small star-wall separations, $z \ll R_{\rm g}$, we replace the
second argument of this function by zero. Moreover, we assert that,
as the dominant contribution to the osmotic pressure for distances
$s < R_{\rm g}$ comes from the first few monomers along the chains
colliding with the wall,
the degree of polymerization $N$ of the chains should be irrelevant
if the chains are long. Hence, all $R_{\rm g}$-dependence 
of the pressure should drop
out, with the implication $h(x,0) \sim x^{-3}$ for $x \ll 1$ and hence
$\Pi(s) \sim s^{-3}$ in this regime.  

Third, we point out that bringing a star 
with $f$ arms at a small distance to a flat
wall, creates a conformation which is very similar to one of an
isolated star with $2f$-arms, as shown in Fig.\ \ref{mirror.fig}.
Hence, it is not surprising that at small star-wall separations,
one recovers for the radial dependence osmotic pressure the scaling laws
pertinent to an isolated star.

Finally, by inserting Eq.\ (\ref{osmotic.eq}) into Eq.\ (\ref{fsw.2})
and carrying out the integration, we find that for small star-wall
distances, $z \ll R_{\rm s}$, the force scales as
$F_{\rm sw}(z) \sim (k_BT)/z$, thus giving rise to a logarithmic
effective star-wall potential
$V_{\rm sw}(z) \sim -k_BT\ln(z/R_{\rm s})$. The latter is indeed
in full agreement with predictions from scaling arguments arising
in polymer theory \cite{pincus:porc:91,Witten:Pincus:86:1,Witten:Pincus:86:2}.
This is a universal
result, in the sense that it also holds for single chains,
be it real or ideal, as it can also be read off from the exact result,
Eq.\ (\ref{erich.1}), using the property ${\rm erf}(x) \sim x$ for $x \to 0$.
Thus, the proposed functional form for the osmotic pressure,
Eq.\ (\ref{osmotic.eq}), combined with Eq.\ (\ref{fsw.2}) for the 
calculation of the effective force, has the following remarkable
property:
it yields the correct result both at small and at large star-wall
distances and therefore appears to be a reliable analytical tool for
the calculation of the effective force at {\it all} star-wall 
distances. At the same time, it contains two free parameters,
$\Lambda$ and $\kappa$ which allow some fine tuning when the
predictions of the theory are to be compared with simulation results,
as we will do below. Yet, we emphasize that this freedom is {\it not}
unlimited: on physical grounds, $\kappa$ must be of the order of 
$R_{\rm g}^{-1}$ and $\Lambda$ must be a number of order unity for
all functionalities $f$, as the dominant, $f^{3/2}$-dependence of the
osmotic pressure prefactor has been already explicitly taken into account
in Eq.\ (\ref{osmotic.eq}).

We are now in a position to write down the full expression for the
star-wall force, by using Eqs.\ (\ref{fsw.2}) and (\ref{osmotic.eq}).
The result reads as:
\begin{equation}
\label{fsw.eq}
\frac{R_{\rm s} F_{\rm sw}(z)}{k_BT} = \Lambda f^{3/2}
\begin{cases}
             \frac{R_{\rm s}}{z} +
                  \frac{z}{R_{\rm s}}(2\xi - 1)
             & \text{for $z \leq R_{\rm s}$};\\
                  2\xi\exp[-\kappa^2(z^2 - R_{\rm s}^2)]
                  & \text{for $z > R_{\rm s}$}.
\end{cases}
\end{equation}
Note the dominant, $\sim 1/z$-dependence for $z \to 0$. Accordingly, 
the effective interaction potential $V_{\rm sw}(z)$ between a star and a 
flat, hard wall held at a center-to-surface distance $z$ from each other
reads as:
\begin{equation}
\label{vsw.eq}
\beta V_{\rm sw}(z) = \Lambda f^{3/2}
\begin{cases}
   -\ln(\frac{z}{R_{\rm s}})-(\frac{z^{2}}
               {R_{\rm s}^{2}}-1)(\xi-\frac{1}{2}) + \zeta
       & \text{for $z\leq R_{\rm s}$};
       \\
       \zeta\,{\rm erfc}(\kappa z)/{\rm erfc}(\kappa R_{\rm s})
       & \text{for $z > R_{\rm s}$},
\end{cases}
\end{equation}
with the inverse temperature $\beta = \left(k_BT\right)^{-1}$, 
the additional constant
\begin{equation}
\zeta=\frac{\sqrt{\pi}\xi}
         {\kappa R_{\rm s}}\exp(\kappa^{2}R_{\rm s}^{2})\,
         {\rm erfc}(\kappa R_{\rm s})
\label{zeta.eq}
\end{equation}
and the complementary error function ${\rm erfc}(x) = 1 - {\rm erf}(x)$.
This completes our theoretical analysis of the star polymer-wall force
and the ensuing effective interaction potential. The comparison with
simulation data and the determination of the free parameters in the
theory will be discussed in section \ref{simulation}. We now proceed
with the calculation of the effective force between a star and a
spherical hard particle, where effects of the colloid curvature
become important.

\subsection{A star polymer and a spherical colloid}

We apply the same idea as for the 
case of the hard wall: 
the effective force acting at the center of
the objects 
is obtained by integrating the osmotic pressure 
exerted by the polymer on the surface of the colloid.
In Fig.\ \ref{star-colloid}, the geometrical situation
is displayed: within the
corona radius of the star polymer $R_{\rm s}=\sigma_{\rm s}/2$, the
osmotic pressure is determined by scaling laws; the outer regime is shadowed
and signifies the Gaussian decay of the osmotic pressure. At center-to-surface
distance $z$ (center-to-center distance $r = z+R_{\rm c}$),
the integration of the osmotic pressure is carried out over
the contact surface between star and colloid. Taking into account the
symmetry of the problem, e.g., 
its independence of the azimuthal angle, we obtain
the force $F_{sc}(z)$ between the star and the colloid as:
\begin{equation}
F_{\rm sc}(z) = 2\pi R_{\rm c}^2 \int_0^{\theta_{\rm max}}
{\rm d}{\theta}\,\sin\theta\,\Pi(s)\cos\vartheta,
\label{fsc.1}
\end{equation}
where $\vartheta$ and $\theta$ are polar angles emanating from the 
center of the star polymer and the colloid, respectively.
The variables $\vartheta$ and $\theta$ can be eliminated in favor of
the variable $s$, which denotes the distance between the center of the star
and an arbitrary point on the surface of the colloid. This elimination
is achieved by taking into consideration the
geometrical relations (see Fig.\ \ref{star-colloid}):
\begin{equation}
s\,\sin\vartheta = R_{\rm c}\sin\theta
\label{geometry.1}
\end{equation}
and
\begin{equation}
s\,\cos\vartheta + R_{\rm c}\cos\theta = R_{\rm c} + z.
\label{geometry.2}
\end{equation}
Eqs.\ (\ref{fsc.1}), (\ref{geometry.1}) and (\ref{geometry.2}) yield for
the star-colloid effective force the transformed integral:
\begin{equation}
\label{Fz.eq}
F_{\rm sc}(z) = \frac{\pi R_{\rm c}}{(z+ R_{\rm c})^{2}} 
 \int_{z}^{s_{\rm max}} {\rm d}s
 \left[(z+ R_{\rm c})^{2} - R_{\rm c}^{2} + s^2\right] \Pi(s)
\end{equation}
The maximum integration distance, $s_{\rm max}$,
depends geometrically on $\theta_{\rm max}$, as well as on
the distance $z$ of the
star polymer to the surface of the colloid and on $R_{\rm c}$.
The relation reads as
\begin{eqnarray}
\nonumber
s_{\rm max} & = & \sqrt{\left[z +R_{\rm c}(1-\cos{\theta_{\rm max}})\right]^2
        + (R_{\rm c}\sin{\theta_{\rm max}})^{2}} 
\\
& = & 
\frac{1}{q}\sqrt{\left[qz +R_{\rm g}(1-\cos{\theta_{\rm max}})\right]^2
        + (R_{\rm g}\sin{\theta_{\rm max}})^{2}}.
\label{rmax.eq}
\end{eqnarray}

By introducing Eq.\ (\ref{osmotic.eq})
into Eq.\ (\ref{Fz.eq}), an analytic expression for the effective force
follows, which reads as
\begin{equation}
\label{fsc.eq}
\frac{F_{\rm sc}(z)}{k_{\rm B}T} =
\frac{\Lambda f^{3/2}R_{\rm c}}{(z+ R_{\rm c})^{2}}   
\begin{cases}
 \left[ (z+ R_{\rm c})^{2} - R_{\rm c}^{2}\right
           ]\left[\frac{1}{2z^{2}}-\frac{1}{2{R_{\rm
           s}}^{2}} + \Psi_{1}(R_{\rm s})\right ]-
   \ln(\frac{z}{R_{\rm s}}) +
           \Psi_{2}(R_{\rm s})
                   & \text{for $z\leq R_{\rm s}$};
           \\
                 \left[ (z+ R_{\rm c})^{2} - R_{\rm c}^{2}\right
           ] \Psi_{1}(z)  +  \Psi_{2}(z)
                   & \text{for $z> R_{\rm s}$}.
\end{cases} 
\end{equation}
Here, the functions $\Psi_1(x)$ and $\Psi_2(x)$ are given by:
\begin{align}
\Psi_{1}(x) & = \frac{\xi}{R_{\rm s}}
     \exp\left(\kappa^2 R_{\rm s}^2\right)
    \left[\frac{1}{x}\exp\left(-\kappa^2 x^2\right)-
     \frac{1}{s_{\rm max}}\exp\left(-\kappa^2 s_{\rm max}^2\right)\right],
\label{psi1}
\\
\intertext{{\rm and}} 
\Psi_{2}(x) & = \frac{\xi}{R_{\rm s}}
    \exp\left(\kappa^2 R_{\rm s}^2\right)
   \left[\frac{\sqrt{\pi}}{\kappa}
     \left[{\rm erf}\left(\kappa s_{\rm max}\right)
    -{\rm erf}\left(\kappa x\right)\right] + 
     x\exp\left(-\kappa^2 x^2\right) - 
      s_{\rm max}\exp\left(-\kappa^{2}s_{\rm max}^2\right)
\right],
\label{psi2}
\end{align}
where $\xi$ is given by Eq.\ (\ref{xi}). 
Note that, for small distances, both regimes of the osmotic pressure 
contribute to the integral,
whereas for larger distances, $z>R_{\rm s}$, only the Gaussian decay
does so. Due to the additional
dependence of $s_{\rm max}$ on the distance $z$, [see Eq.\ (\ref{rmax.eq})],
an analytical expression for the effective potential $V_{\rm sc}(z)$,
analogous to Eq.\ (\ref{vsw.eq}) for the flat-wall case,
is not possible here. 
 
Some remarks regarding $F_{\rm sc}(z)$ are necessary. First, for small
separations $z$, the force scales as $F_{\rm sc}(z) \sim (k_BT)/z$, the
same behavior found for the flat-wall case. Once more, we obtain the universal
result mentioned above, which has been shown to be also valid
for an ideal chain whose one end is held at a distance $z$
from the surface of a hard sphere. Indeed,
for this case the force is given by the exact relation \cite{eisenriegler:pre:97}: 
\begin{equation}
F_{\rm sc}^{\rm (id)}(z) = k_BT\frac{\partial}{\partial z}\ln
                \left[1 - \left(\frac{R_{\rm c}}{z + R_{\rm c}}\right)
                      {\rm erfc}\left(\frac{z}{L}\right)\right],
\label{erich.3}
\end{equation}
with $L$ being a length scale of order $R_{\rm g}$. Eq.\ (\ref{erich.3})
above, yields $F_{\rm sc}^{\rm (id)}(z) \sim (k_BT)/z$ for $z \to 0$.

Second, let us consider the limit of small 
size ratios $q=R_{\rm g}/R_{\rm c}$. As can be seen
from Eq.\ (\ref{rmax.eq}), the upper integration limit
$s_{\rm max}$ scales as $R_{\rm g}/q$, whereas the decay parameter
$\kappa$ is of the order $R_{\rm g}^{-1}$. 
Hence, $\kappa s_{\rm max} \sim q^{-1}$,
with the implication that for small enough $q$'s, the argument 
$\kappa s_{\rm max}$ in the error function and in the Gaussian in 
Eqs.\ (\ref{psi1}) and (\ref{psi2}) can be replaced by infinity.
As we will shortly see, this is an excellent approximation up to
$q \lesssim 0.3$, as both ${\rm erf}(x)$ 
and $\exp(-x^2)$ approach their 
asymptotic values for $x \to \infty$ rapidly. Then, the implicit
$z$-dependence of the force $F_{\rm sc}(z)$ through $s_{\rm max}$ 
drops out and a $z$-integration of the latter can be analytically
carried out to obtain an effective star polymer-colloid potential
$V_{\rm sc}^{\infty}(z)$ which reads as \cite{joe:likos}:
\begin{equation}
\beta V_{\rm sc}^{\infty}(z) = \Lambda f^{3/2}
                  \left(\frac{R_{\rm c}}{z+R_{\rm c}}\right)
\begin{cases}
   -\ln(\frac{z}{R_{\rm s}})-(\frac{z^{2}}
                 {R_{\rm s}^{2}}-1)(\xi-\frac{1}{2}) + \zeta
       & \text{for $z\leq R_{\rm s}$};
       \\
       \zeta\,{\rm erfc}(\kappa z)/{\rm erfc}(\kappa R_{\rm s})
       & \text{for $z > R_{\rm s}$},
\end{cases}
\label{vsc.eq}
\end{equation} 
with the constant $\zeta$ given by Eq.\ (\ref{zeta.eq}).
Clearly, in the limit $R_{\rm c} \to \infty$ ($q \to 0$), 
corresponding to a flat wall, Eq.\ (\ref{vsc.eq}) reduces to the 
previously derived result, Eq.\ (\ref{vsw.eq}). It is a remarkable
feature that all effects of curvature are taken into account by
the simple geometrical prefactor $R_{\rm c}/(z+R_{\rm c})$, for
sufficently small size ratios $q$. In this respect, the above result
bears close similarity to the well-known Derjaguin approximation \cite{hunter}.

\section{Simulation}
\label{simulation}

\subsection{The simulation model}

In order to check the theoretical prediction of the forces at hard
objects, we performed a monomer-resolved Molecular Dynamics (MD)
simulation and calculated the mean force at the center of the star
polymer to compare the data with theory.
The model is based on the ideas of simulation methods applied on linear
polymers and on a single star \cite{Grest:87:1,Grest:94:1}. 
The main features are as follows.

A purely repulsive and truncated Lennard-Jones potential acts
between all $Nf$ monomers at distances $r$:
\begin{eqnarray}  
V_{\rm LJ}(r) = 
\begin{cases}
                        4\epsilon\left[
                        \left(\frac{\sigma_{\rm LJ}}{r}\right)^{12}
                        -\left(\frac{\sigma_{\rm LJ}}{r}\right)^{6}
                        +\frac{1}{4}
                     \right] 
                   & \text{for $r\leq 2^{1/6}\sigma_{\rm LJ}$};
             \\
                   0 
                   & \text{for $r>2^{1/6}\sigma_{\rm LJ}$}.
\end{cases}
\label{vlj}
\end{eqnarray}
Here, $\sigma_{\rm LJ}$ is the microscopic length scale
of the beads and $\epsilon$ sets the energy scale. In accordance 
with previous work \cite{jusufi:macromolecules:99}, we have chosen
$T = 1.2\epsilon/k_B$. 

An attractive FENE (finite extensible nonlinear
  elastic) potential additionally acts between
  neighboring monomers along a chain \cite{Grest:87:1}:
\begin{eqnarray}
V_{\rm FENE} (r)= 
\begin{cases}
            -15\epsilon\left(\frac{R_{0}}{\sigma_{\rm LJ}}\right)^{2}
                   \ln\left[1-\left(\frac{r}{R_{0}}\right)^{2}\right]
                   & \text{for $r\leq R_{0}$};
             \\
                   \infty 
                   & \text{for $r>R_{0}$}.
\end{cases} 
\label{vfene}
\end{eqnarray}
This interaction diverges at $r=R_0$, which determines the maximal 
relative displacement of two neighboring beads. The energy 
$\epsilon$ is the same as in Eq.\ (\ref{vlj}), whereas for the 
length scale $R_0$ we have chosen the value $R_0 = 1.5\,\sigma_{\rm LJ}$. 

To accommodate the polymer arms, 
a hard core with radius $R_{\rm d}$ is introduced at the
center of the star; its size depends on the arm number $f$. 
Accordingly, the interactions between the monomers and
the central particle were introduced. All monomers had a repulsive interaction
$V_{\rm LJ}^p(r)$
of the truncated and shifted Lennard-Jones type with the central
particle,
\begin{equation}
V_{\rm LJ}^p(r) =
\begin{cases}
   \infty & \text{for $r \leq R_{\rm d}$};\\
   V_{\rm LJ}(r - R_{\rm d}) & \text{for $r > R_{\rm d}$},
\end{cases}
\label{mon.core1}
\end{equation}
whereas the innermost monomers in the chain an additional
attractive potential $V_{\rm FENE}^p(r)$ of the FENE type, namely
\begin{equation}
V_{\rm FENE}^p(r) =
\begin{cases}
   \infty & \text{for $r \leq R_{\rm d}$};\\
   V_{\rm FENE}(r - R_{\rm d}) & \text{for $r > R_{\rm d}$}.
\end{cases}
\label{mon.core2}
\end{equation}
Finally,
all monomers interact with the colloid or with the wall by a
hard potential.
We note that exactly this simulation model was already used by
Grest {\it et al.}\ in their simulations of linear and star 
polymers in good solvent
conditions \cite{Grest:87:1,Grest:94:1}.

The timestep is typically $\Delta t=0.002t^{*}$ with
$t^{*}=\sqrt{m\sigma_{\rm LJ}^{2}/\epsilon }$ being the associated
time unit and $m$ the monomer mass.  
After a long equilibration time ($500\,000$ MD steps), 
the mean force at the core of the 
star whose center is held at the position ${\bf R}$
and its dependence on the arm number $f$
separations, is 
calculated as the expectation value over all instantaneous forces
acting on the star core:
\begin{equation}
{\bf F}({\bf R}) = \left < -\nabla_{{\bf R}} \left [
              \sum_{k=1}^{fN} V_{\rm LJ}^p(|{\bf r}_k - {\bf R}|) +
              \sum_{l=1}^{f} V_{\rm FENE}^p(|{\bf r}_l - {\bf R}|)
              \right]\right>,
\label{force.jusufi}
\end{equation}
where the first sum is carried over all $fN$ monomers of the star
and the second only over the $f$ innermost monomers of its chains.
The direct force between the central particle
and the wall did not need to be considered,
as the center-to-surface distance was always
kept at values where this force was vanishingly small.
Note that choosing the origin of the coordinate system on the
surface of the colloidal particle or wall, at the 
point of nearest separation between the star center and this surface,
and also the $z$-axis in the direction connecting this origin
with the star center, we immediately obtain $R \equiv |{\bf R}| = z$.

We have carried out simulations for a variety of arm numbers $f$ and
size ratios $q$, allowing us to make systematic predictions for the
$f$- and $q$-dependencies of all theoretical parameters.
In attempting to compare the simulation results with the theoretical
predictions, one last obstacle must be removed: in theory, the 
fundamental length scale characterizing the star is the corona radius
$R_{\rm s}$. The latter, however, is not directly measurable in a 
simulation in which, instead, we can only assess to the radius
of gyration $R_{\rm g}$. Yet, we have previously found 
that the ratio between the two
remains fixed for all considered arm numbers
$f$, having the value
$R_{\rm s}/R_{\rm g} \simeq 0.66$ \cite{jusufi:macromolecules:99}.
We now proceed with the presentation of our MD results.
 
\subsection{Star-wall and star-colloid interactions}

We consider at first a star polymer near a hard wall. The theoretical
prediction of the effective interaction force is given in 
Eq.\ (\ref{fsw.eq}). First, we consider the limit of small
separations, $z \to 0$, which allows us on the one hand to test
the theoretical prediction $F_{\rm sw}(z) \cong k_BT\Lambda f^{3/2}/z$ 
there and on the other hand to fix the value of the prefactor $\Lambda$,
which is expected to have in general a weak $f$-dependence. 
For this prefactor, some semi-quantitative theoretical predictions 
already exist: For $f=1,2$ the prefactor may be calculated from the bulk and
the  ordinary surface critical exponents $\nu,\gamma$ and
$\gamma^o,\gamma^o_1$ of the $n$-vector model.
For $n=0$ this results in
$\Lambda(f=1)=(\gamma-\gamma^o_1)/\nu$ and
$2^{3/2}\Lambda(f=2)=(\gamma-\gamma^o)/\nu=1/\nu$
\cite{Dietrich:1981,Ohno:1988}.
Numerical values for the exponents are known from renormalization group
theory
and simulation \cite{Diehl:1998,Hegger:1994}
and yield $\Lambda(f=1)\approx 0.83$ and $\Lambda(f=2)\approx 0.60$.
On the other hand, for very large functionalities, $f \gg 1$, one can
make an analogy between a star at distance $z$ from a wall and two
star polymers whose centers are kept at distance $r = 2z$ from each
other \cite{pincus:porc:91}. Indeed, for very large $f$, the conformations
assumed by two stars brought close to each other is one in which 
the chains of each star retract to the half-space where the center
of the star lies, a situation very similar to the star-wall case.
Then, one can make the approximation $F_{\rm sw}(z) \cong F_{\rm ss}(2z)$,
where $F_{\rm ss}$ denotes the star-star force. For the latter, it is
known \cite{likos:etal:prl:98} that it has the form:
\begin{equation}
F_{\rm ss}(r) = \frac{5}{18}f^{3/2}\frac{1}{r}\qquad\qquad(r \to 0),
\end{equation}
implying for the coefficient $\Lambda$ the asymptotic behavior:
\begin{equation}
\lim_{f \to \infty}\Lambda(f) \equiv \Lambda_{\infty} = 
\frac{5}{36} \cong 0.14.
\label{lambda.asympt}
\end{equation} 

Since there is no theory concerning the values of $\Lambda$ in the
intermediate regime of $f$, $\Lambda$ is used as fit parameter.
Its value can be obtained by plotting the inverse force 
$1/F_{\rm sw}(z)$ against $z$
for small separations $z$ to the hard wall. The results are shown in 
Fig.\ \ref{inverse-force}. Looking first at the inset, 
we see that, as for the earlier case of
star-star interactions \cite{jusufi:macromolecules:99}, the reciprocal
force curves do not go through the origin, as a result of the finite
core size, $R_{\rm d}$. Once this is subtracted, though, straight
lines passing through the origin are obtained, verifying in this
way the $1/z$-behavior of the force and the associated logarithmic
dependence of the effective potential at small separations. The
values for $\Lambda(f)$ can be immediately read off from the 
slope of the curves and they are summarized in Table
\ref{TABparameters}.
There and in Fig.\ \ref{lk} 
we see that $\Lambda$ is indeed a decreasing function of $f$ 
but the asymptotic value
$\Lambda_{\infty}=5/36$ is still not achieved at arm numbers as high
as $f = 100$. 

The decay parameter $\kappa$ is fixed by looking at the force at 
larger separations and the obtained are also summarized in 
Table \ref{TABparameters} and shown in Fig.\ \ref{lk}. As expected,
$\kappa$ is of the order $R_{\rm g}^{-1}$, as witnessed by the
fact that the product $\kappa R_{\rm s}$
is of order unity. A monotonic increase of $\kappa\sigma_{\rm g}$
with the arm number $f$ is observed, consistent with the view that
for large $f$ stars form compact objects with an increasingly small
diffuse layer beyond their coronae \cite{likos:etal:prl:98}.

With parameters $\Lambda$ and $\kappa$ {\it once and for all fixed}
from the star-wall case, we now turn our attention to the 
interaction of a star polymer at a hard
sphere of finite radius $R_{\rm c}$, equivalently size ratios 
$q \ne 0$. Here, the force is given by the full expressions
of Eqs.\ (\ref{fsc.eq}), (\ref{psi1}) and (\ref{psi2}); for small
enough size ratios $q$, the approximation $\kappa s_{\max} \to \infty$
gives rise to a simplified expression for the force and to the
analytical formula, Eq.\ (\ref{vsc.eq}) for the effective star-colloid
potential. Our purpose is twofold: to test the validity of these
simplified expressions as a function of $q$ and also to find 
an economical way to parameterize $s_{\rm max}$ as a function of
$q$ for those values of the size ratio for which the approximation
$\kappa s_{\max} \to \infty$ turns out to be unsatisfactory.

We show representative results for fixed arm number $f = 18$ and
varying $q$ in Fig.\ \ref{star-coll_18}; results for different 
$f$-values are similar. It can be seen that the simplified 
result arising from allowing $s_{\rm max} \to \infty$ yields
excellent results up to size ratios $q \lesssim 0.3$,
see Figs.\ \ref{star-coll_18}(a) and (b). However,
above this value, the approximation of integrating the osmotic
pressure up to infinitely large distances breaks down, as it
produces effective forces that are larger than the simulation
results, especially at distances $z$ of order of the radius of
gyration $R_{\rm g}$. These are the dashed lines shown in 
Figs.\ \ref{star-coll_18}(c)-(e). The overestimation of the force
is not surprising: as can be seen from
Fig.\ \ref{star-colloid} and Eq.\ (\ref{Fz.eq}), we are integrating
a positive quantity beyond the physically allowed limits and this
will inadvertently enhance the resulting force. Hence, we have to
impose a finite upper limit $s_{\rm max}$ for size ratios 
$q > 0.3$ in order to truncate the contribution of the Gaussian
tail in the integral of the osmotic pressure in Eq.\ (\ref{Fz.eq}).
 
In Fig.\ \ref{pov} a typical snapshot of a star polymer at a colloid
illustrates the situation. One can see that the main contribution of
the osmotic pressure results from in the inner region of the star. The
outer region of the chains only interact weakly with the sphere.
The question now is how the value of $s_{\rm max}$ must be chosen.
As can be seen from Eq.\ (\ref{rmax.eq}), this quantity is dependent
on, $R_{\rm s}$, $z$ and $\theta_{\rm max}$. (The latter depending
on $q$ means that $\theta_{\rm max}$ and $q$ should not be treated as
independent quantities.) It would be indeed
most inconvenient if for every combination of these we would have
to choose a different upper integration limit. Hence, we have 
attempted to transfer all dependence of $s_{\rm max}$ onto the
maximum integration angle $\theta_{\rm max}$. We found that this
is indeed possible and, in fact, the angle $\theta_{\rm max}(q)$
has a very weak $q$-dependence: starting with a value
$\theta_{\rm max} \approx 45^{\rm o}$ at $q = 0.3$, we find that
it then quickly saturates into the value $\theta_{\rm max}\approx 30^{\rm o}$ 
for all $q \gtrsim 0.35$. In this way, we are able to obtain the
corrected curves denoted by the solid lines in 
Figs.\ \ref{star-coll_18}(c)-(e), showing excellent agreement with
the simulation results.

We finally turn our attention to the $f$-dependence of the forces
for a fixed value of the size ratio, $q = 0.33$. In 
Fig.\ \ref{star-coll_0.3} we show the simulation results compared
with theory for a wide range of arm numbers, $5 \leq f \leq 50$.
For the theoretical fits, the values of $\Lambda$ and $\kappa$
from Table \ref{TABparameters} were used, whereas the value of
the maximum integration angle was kept fixed at 
$\theta_{\rm max} = 30^{\rm o}$ for all $f$-values. The agreement
between theory and simulation is very satisfactory. 

Thus, our conclusions for the star polymer-colloid interaction
read as follows: the general, analytical expression for the
{\it force} between the two is given by Eqs.\ (\ref{fsc.eq}),
(\ref{psi1}) and (\ref{psi2}), supplemented by Eq.\ (\ref{rmax.eq})
in which the angle $\theta_{\rm max}$ has to be chosen as 
discussed above for $q \gtrsim 0.3$. An analytical formula for
the effective {\it interaction potential} $V_{\rm sc}(z)$ is not
possible for such size ratios. Rather, the results for the effective
force have to be integrated numerically in order to obtain 
$V_{\rm sc}(z)$.
For size ratios $q \lesssim 0.3$
on the other hand, the approximation $s_{\rm max} \to \infty$
in Eqs.\ (\ref{fsc.eq}), (\ref{psi1}) and (\ref{psi2}) for the 
effective force can be made, thereby also allowing us to derive
a simple, accurate, and analytic form for the interaction potential
between a star polymer and a colloid, given by Eq.\ (\ref{vsc.eq}).
These results form the basis of  
the statistical-mechanical treatment of
star polymer-colloid mixtures in terms of standard liquid-state
theories; the availability of analytical results for
the pair interactions greatly facilitates the latter.
A many-body theory of star polymer-colloid mixtures was put forward 
recently by 
Dzubiella {\it et al.}\ \cite{joe:likos}, who employed the 
above-mentioned effective interactions in order to study the fluid-fluid
separation (demixing transition) in such systems. The very good
agreement with experimental results obtained in that work offers
further corroboration of the validity of the interactions presented 
here. 

As the ultimate goal of the derivation of the interactions we present
here is precisely to allow theoretical investigations of star polymer-colloid
mixtures, we present in the next section a short account of 
a revision of the star-star interactions for the case of very low
arm numbers. In this way, mixtures containing stars with arbitrary 
arm numbers, ranging from free chains ($f = 1, 2$) to the 
``colloidal limit'' of $f \gg 1$ can be studied in full generality.

\section{Revised star-star interaction for small arm numbers}
\label{revised}

The effective interaction between two stars in a good solvent was
recently derived by theoretical scaling arguments and verified by
neutron scattering and molecular simulation \cite{likos:etal:prl:98,stellbrink:ecis98,stellbrink:ecis00,jusufi:macromolecules:99}, 
leading thereafter to the phase diagram of the system \cite{watzlawek:etal:prl:99,watzlawek:etal:jpcm:98}.
The pair potential was modeled by an
ultrasoft interaction which is logarithmic for an inner core and
shows a Yukawa-type exponential decay at larger
distances \cite{likos:etal:prl:98,watzlawek:etal:prl:99}:
\begin{equation}
V_{\rm ss}(r)  = \frac{5}{18}k_{\rm B}Tf^{3/2}
\begin{cases}
   -\ln(\frac{r}{\sigma_{\rm s}}) + \frac{1}{1+\sqrt{f}/2}
   & \text{for $r \leq \sigma_{\rm s}$};
   \\
   {\frac{\sigma_{\rm s}/r}{1+\sqrt{f}/2}
    \exp(-\frac{\sqrt{f}}{2\sigma_{\rm s}}(r-\sigma_{\rm s})) }
   & \text{for $r > \sigma_{\rm s}$},
\end{cases}
\label{pot_ss}
\end{equation}
However,
the theoretical approach giving rise to
Eq.\ (\ref{pot_ss}) does not hold for arm numbers $f\lesssim 10$,
because the
Daoud-Cotton model of a star \cite{Daoud:Cotton:82:1}, on which the Yukawa
decay rests, is not valid for small $f$. In these cases, the interaction
has to a shorter-ranged decay for $r > \sigma_{\rm s}$. The shortcomings
of the blob model can be made evident if one considers the extreme limit
$f = 1$, corresponding to free chains. There, the geometrical blob picture
and the associated ``cone approximation'' \cite{Ohno:89:1} break down.
It is therefore instructive to consider known results about the 
effective interactions between free chains in order to obtain some
insight for the case at hand.  

Most of the work done on chain-chain interactions concerns the effective
potential between the centers of mass of the chains \cite{ard:peter:00,bolhuis:jcp:00,grosberg:82,schaefer:baumgaertner:86,krueger:etal:89}. 
Theoretical approaches considering two chains \cite{krueger:etal:89},
simulations of two chains \cite{grosberg:82,schaefer:baumgaertner:86},
as well as recent, state-of-the-art simulations of many-chain 
systems \cite{ard:peter:00,bolhuis:jcp:00} all reach the conclusion that
the effective center-of-mass to center-of-mass interaction has a Gaussian
form with its range set by the radius of gyration of the chains. 
Here, we are interested in a slightly different interaction, namely
that between the end-monomer of one chain and the end-monomer of the
other. However, at distances of the order of $R_{\rm g}$ or larger,
whether the centers of mass or the end-monomers choice of the two
chains are held fixed should not make much difference. Therefore,
we assume a Gaussian decay of the star-star potential for small
$f$-values and center-to-center distances larger than $\sigma_{\rm s}$.
We emphasize that {\it only} the
large distance decay of the interaction is affected; its form
at close approaches
has to remain  logarithmic \cite{Witten:Pincus:86:2}. Accordingly, we
propose the following star-star pair potential for arm numbers
$f < 10$,
replacing the Yukawa by a Gaussian decay:
\begin{equation}
V_{\rm ss}(r)  = \frac{5}{18}k_{\rm B}Tf^{3/2}
 \begin{cases}
   -\ln(\frac{r}{\sigma_{\rm s}})+\frac{1}{2\tau^{2}\sigma_{\rm s}^{2}} &
   \text{for $r \leq \sigma_{\rm s}$};
 \\
   {\frac{1}{2\tau^{2}\sigma_{\rm
 s}^{2}}\exp\left(-\tau^{2}(r^{2}-\sigma_{\rm s}^{2})\right) } &
   \text{for $r > \sigma_{\rm s}$},
 \end{cases}
\label{pot_ss2}
\end{equation}
where $\tau(f)$ is a free parameter of the order of  $1/R_{\rm g}$ and is
obtained by fitting to computer simulation results, see
Fig.\ \ref{2stars-inset}. For $f = 2$ we obtain the value $\tau = 1.03$
which, together with the potential in Eq.\ (\ref{pot_ss2}) above 
yields for the 
second virial coefficient of polymer solutions
the value $B_2/R_{\rm g}^3 = 5.59$, in agreement with the estimate
$5.5 < B_2/R_{\rm g}^3 < 5.9$ from renormalization group
and simulations \cite{bolhuis:jcp:00}. For $f=5$ we find $\tau = 1.12$,
which leads to $B_2/R_{\rm g}^3 = 11.48$, in accordance with Monte
Carlo simulation results \cite{Ohno:96:1, Rubio:Freire:99:1}. 

The very good agreement between the logarithmic-Gauss-potential
of Eq.\ (\ref{pot_ss2}) and the simulation data for $f < 10$ can be seen in
Fig.\ \ref{2stars-inset}. In the inset of this Figure, it can also be seen
that the Yukawa decay is way too slow
there. Hence, the potential of Eq.\ (\ref{pot_ss}), has a longer
range than the true interaction for small $f$, a property that
explains the discrepancies between the simulated and theoretical
second virial coefficients based on this potential, which have
been reported by Rubio and Freire \cite{Rubio:Freire:99:1} in their
numerical study of low-functionality stars. 
At the same time, with increasing $f$, the roles of
the Gaussian- and Yukawa-decays are reversed: 
in Fig.\ \ref{2stars-f10}, we show simulation and
theory results for $f = 10$. The original, logarithmic-Yukawa 
potential brings about better agreement now, as already established
by earlier studies on stars with high arm 
numbers \cite{jusufi:macromolecules:99,likos:etal:prl:98,stellbrink:ecis98}.
To summarize, we propose two analytic expressions for the effective
star-star potential, valid in complementary regimes of the 
functionality $f$. 
The first one concerns the regime $f \lesssim 10$
with the validity of the logarithmic-Gauss-potential 
of Eq.\ (\ref{pot_ss2}) being established; in the second
regime, $f \ge 10$, the
logarithmic-Yukawa-potential of Eq.\ (\ref{pot_ss}) holds.
We remark that the ultimative decay of the effective interaction for very
long distances is still Gaussian even for very large $f$, but this is not
relevant for $B_{2}$ as it occurs for much larger distances than the corona
diameter.

\section{Summary and concluding remarks}
\label{summary}

In summary, we have presented analytic results for the force
between a colloid and a star polymer in a good solvent, accompanied
with an analytic expression for the corresponding pair potential
which is valid for size ratios $q \lesssim 0.3$. The validity of
these expressions was established by direct comparison with
Molecular Dynamics simulations. It should be noted that our theoretical
approach is in principle generalizable to arbitrary geometrical
shapes for the hard particle, thus opening up the possibility 
for studying effective forces between stars and hard ellipsoids,
platelets etc. Further, a revised form for the star-star interaction
for small functionalities has been presented, while at the same time
the logarithmic-Yukawa form of this interaction remains valid for
functionalities $f \gtrsim 10$.

The practical advantage of the present results is that they 
greatly facilitate
the study of structural and thermodynamic properties of 
concentrated star polymer-colloid mixtures; a first attempt towards
this has already been undertaken \cite{joe:likos}.
In this work, we limited ourselves to the case
where the star is smaller than the colloid, i.e., $q < 1$. 
The study of the inverse case may be possible by applying the 
ideas presented here, however additional complications arise  
through the possibility of the star to ``surround'' the smaller,
colloidal particle, in which case one part of the arms acts to 
bring about a repulsion with the colloid and another causes an
effective attraction between the two. 
Furthermore, a pair potential picture for the many-body system become
more and more questionable for larger
$q$ as effective many-body forces \cite{Ferber:Jusufi:99:1}
will play a more dominant role in this case.

\acknowledgments 
We thank Prof.\ E. Eisenriegler and Dr.\ Martin Watzlawek for helpful discussions.
This work has been supported by the Deutsche Forschungsgemeinschaft 
within the SFB 237.

\clearpage 
\bibliography
{/home/jusufi/paper/bib/general,/home/jusufi/paper/bib/polymer,/home/jusufi/paper/bib/depletion,/home/jusufi/paper/bib/stars,/home/jusufi/paper/bib/future}
\clearpage

\begin{table}
\begin{center}
\begin{tabular}{|c|c|c|c|}\hline 
$\qquad f\qquad$ & $\qquad R_{\rm d}/R_{\rm s}\qquad$ & 
$\qquad \Lambda \qquad$ & $\qquad \kappa R_{\rm s} \qquad$  \\ \hline \hline
2 & 0.006& 0.46 & 0.58\\
5 & 0.018 & 0.35 & 0.68\\
10 & 0.06 & 0.30 & 0.74\\
15 & 0.12 & 0.28 & 0.76\\
18 & 0.09 & 0.27 & 0.77\\
30 & 0.12 & 0.24 & 0.83\\
40 & 0.152 & 0.24 & 0.85\\
50 & 0.152 & 0.23 & 0.86\\
80 & 0.273 & 0.22 & 0.88\\
100 & 0.303 & 0.22 & 0.89\\
\hline
\end{tabular}
\end{center}
\caption{The fit parameters arising from the comparison between
  theory and simulation for the star-wall and star-colloid interaction. 
  The values of
  $R_{\rm d}$ shown here are not exactly the same as the input core
  size; they are just in the same order of magnitude, deviating only slightly
  from the real input value. They are still corresponding to microscopic
  length, and are thus irrelevant at length scales $r \sim \sigma_{\rm s}$.
  $\Lambda$ is the overall prefactor and $\kappa$ the inverse Gaussian decay
  length, both used in Eqs.\ (\ref{fsc.eq}) and (\ref{vsc.eq}). 
  $\sigma_{\rm s} = 2R_{\rm s} = 0.66\,\sigma_{\rm g}$ 
  denotes the corona diameter of the stars,
  as measured during the simulation.}
\label{TABparameters}
\end{table}
%\clearpage

\begin{figure}
\begin{center}
\includegraphics[width=15.0cm]{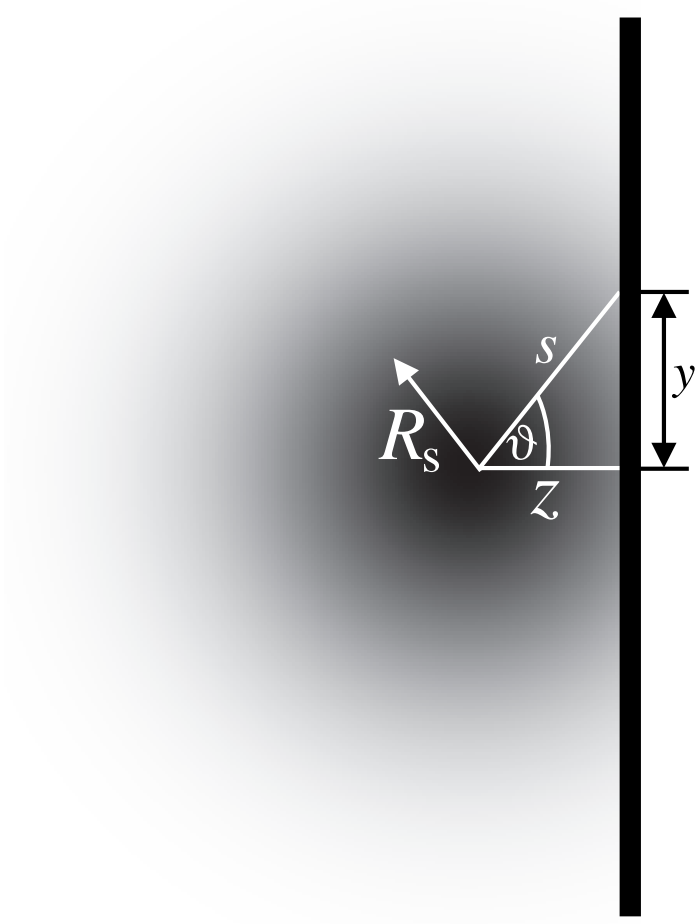}
\end{center}
\caption{Star polymer (black-shadowed particle) interacting with a
  flat wall. The star
  polymer consists of a inner core region, where the scaling
  behavior is dominant, whereas the outer regime is shadowed and
  indicates the exponential decay of the osmotic pressure.
 }
\label{star-wall}
\end{figure}

\begin{figure}
\begin{center}
\includegraphics[width=12.0cm,angle=-90]{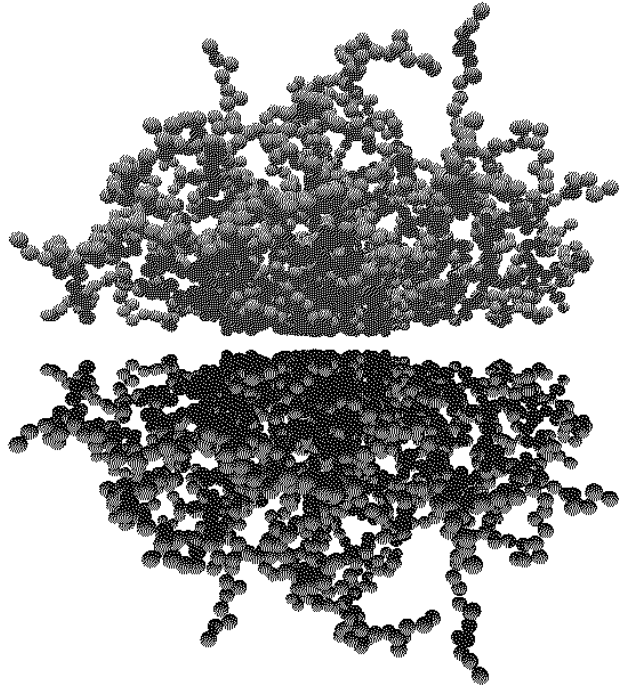}
\end{center}
\caption{Snapshot of a simulation
 showing a star polymer interacting with a flat wall, 
 at a small center-to-surface distance. The mirror-reflected image of
 the star, on the right, helps demonstrate that the configuration
 is similar to that of an isolated star with twice as many arms.}
\label{mirror.fig}
\end{figure}

\begin{figure}
\begin{center}
\includegraphics[width=15.0cm]{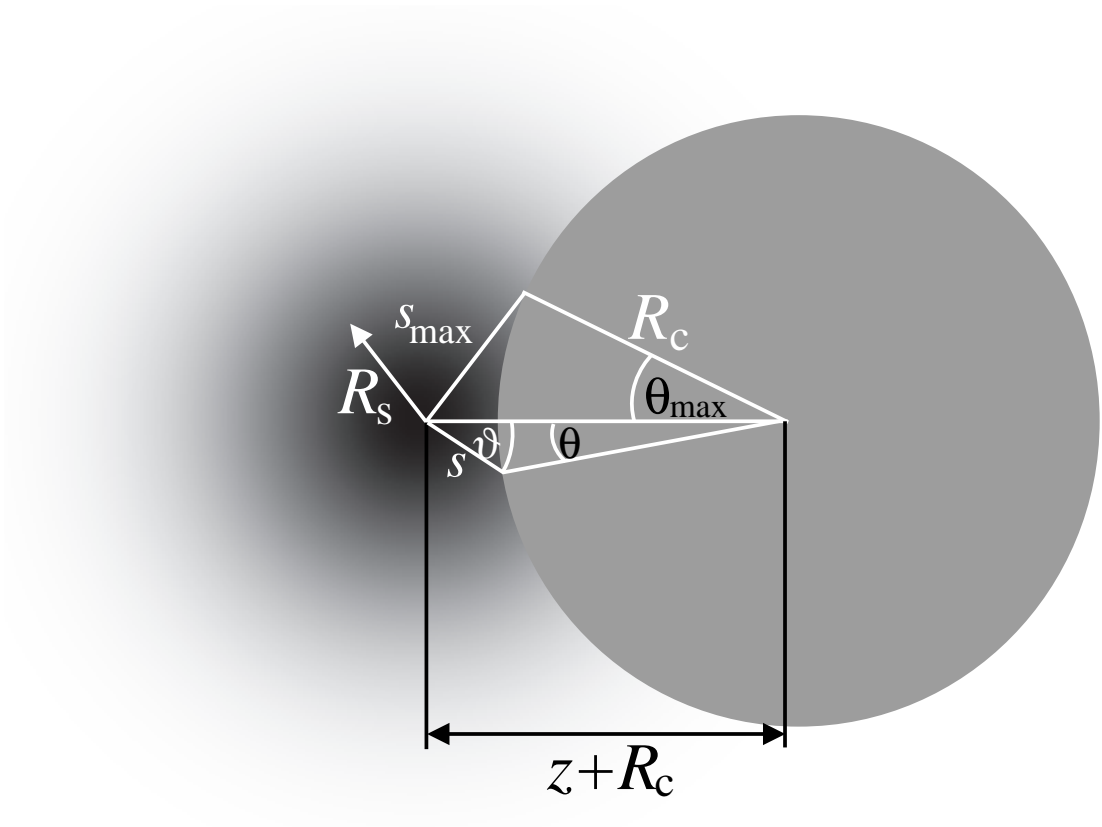}
\end{center}
\caption{Star polymer (black-shadowed particle) interacting with a
  colloidal particle (grey sphere). The dark and shadowed regions 
  of the star have the same meaning as in Fig.\ \ref{star-wall}.
 }
\label{star-colloid}
\end{figure}

\begin{figure}
\begin{center}
\includegraphics[width=12.0cm,angle=-90]{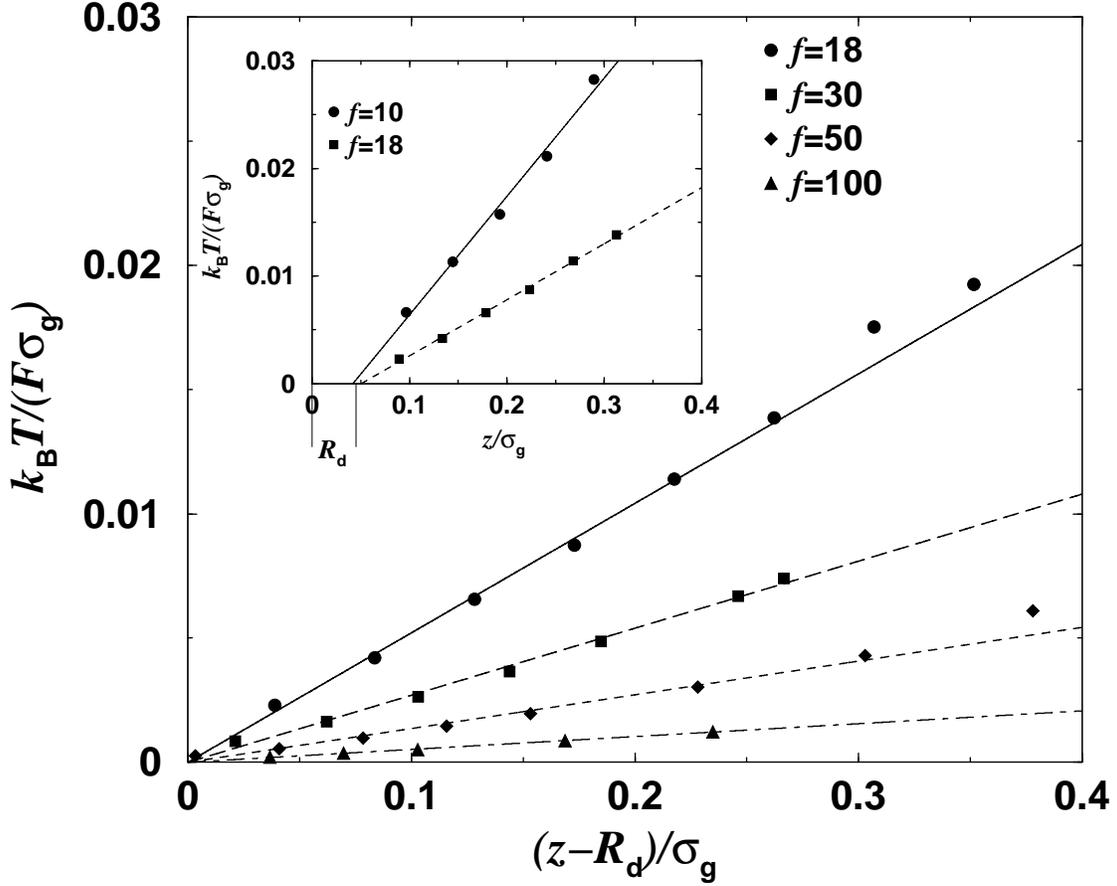}
\end{center}
\caption{Reciprocal effective force between a star polymer and a hard
 flat wall plotted against the distance $z$ between
 the star center to the surface of the wall for small
 $z$-values. The dependence $F(z) \sim 1/z$
 is confirmed by the simulation results (symbols). The prefactor of
 the potential depends on $f$ and manifests itself in the different 
 slopes of the
 reciprocal forces. The inserted plot shows the divergence of the
 force at the distance $z=R_{\rm d}$, which is subtracted from $z$ 
 in the outset of the plot, to achieve divergence of the force in the 
 origin.   
 }
\label{inverse-force}
\end{figure}

\begin{figure}
\begin{center}
\includegraphics[width=12.0cm,angle=-90]{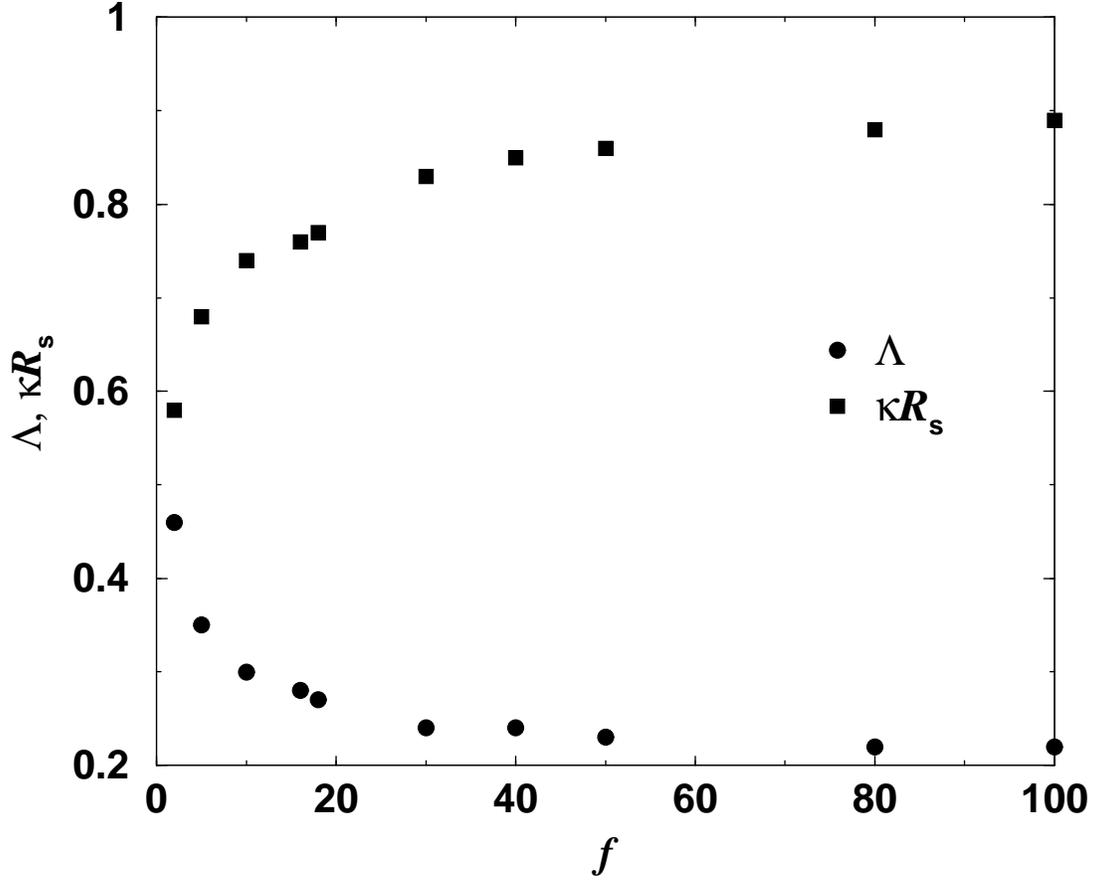}
\end{center}
\caption{The prefactor of $\Lambda$ 
and the decay parameter
$\kappa$ of Eq.\ (\ref{fsw.eq}) plotted against
the functionality $f$.
The value of $\Lambda = 5/36\approx 0.14$ for $f\gg 1$ is not reached
but the simulation data tend to this value very slowly.
$\kappa$ shows a monotonic increase with arm number $f$.}
\label{lk}
\end{figure}

\begin{figure}
\begin{center}
\includegraphics[width=10.0cm,angle=-90]{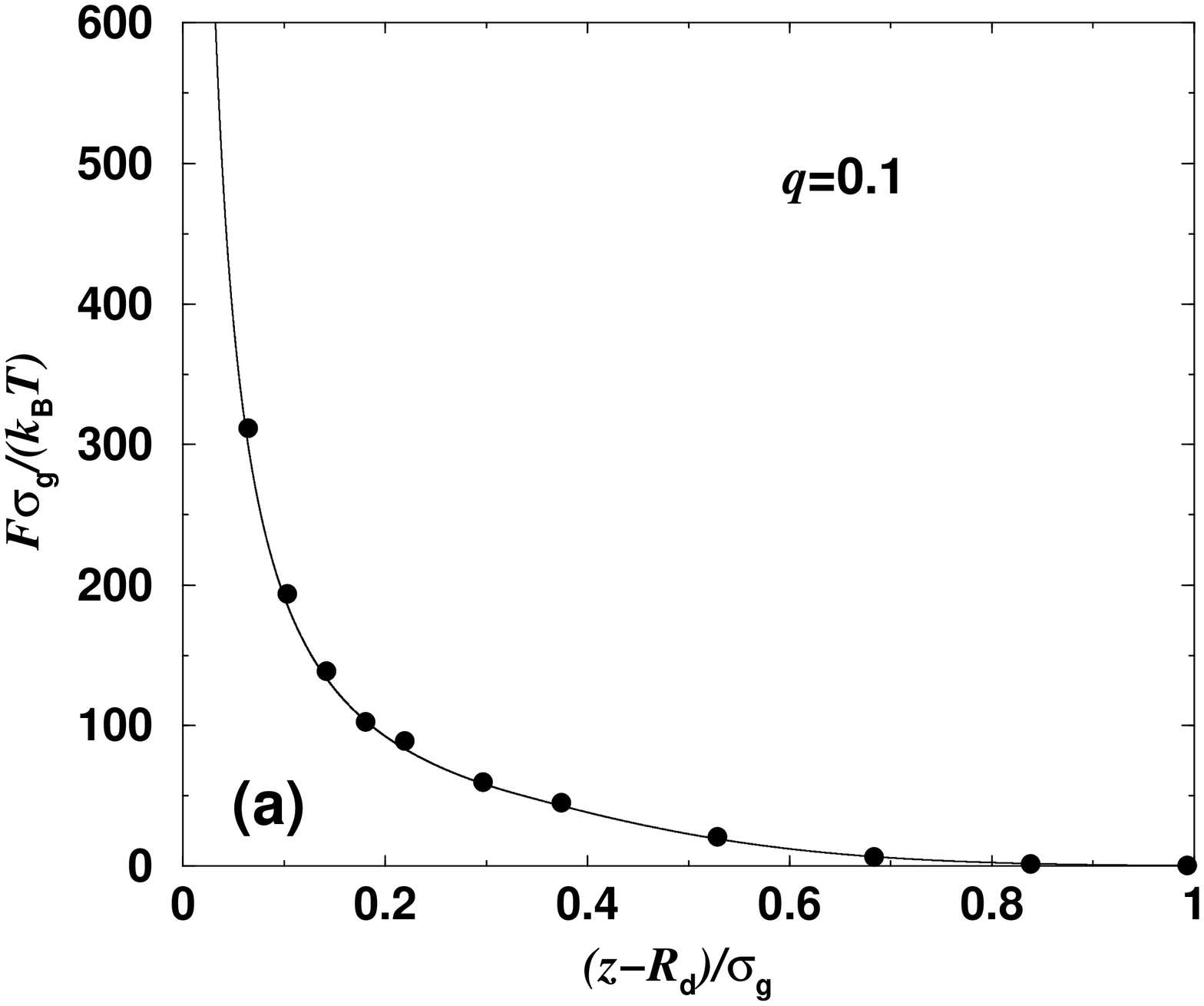}
\includegraphics[width=10.0cm,angle=-90]{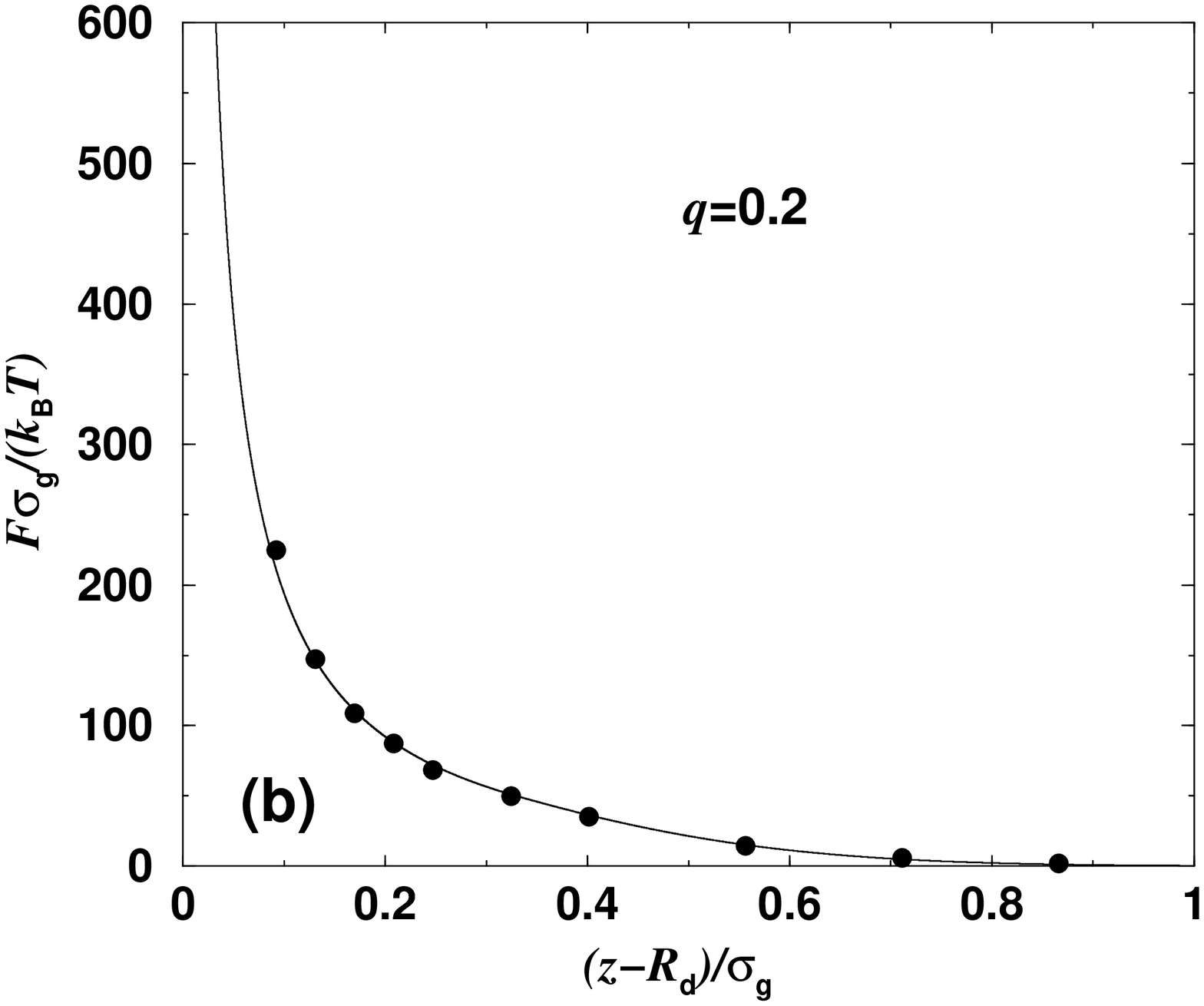}
\end{center}
\end{figure}
\clearpage

\begin{figure}
\begin{center}
\includegraphics[width=10.0cm,angle=-90]{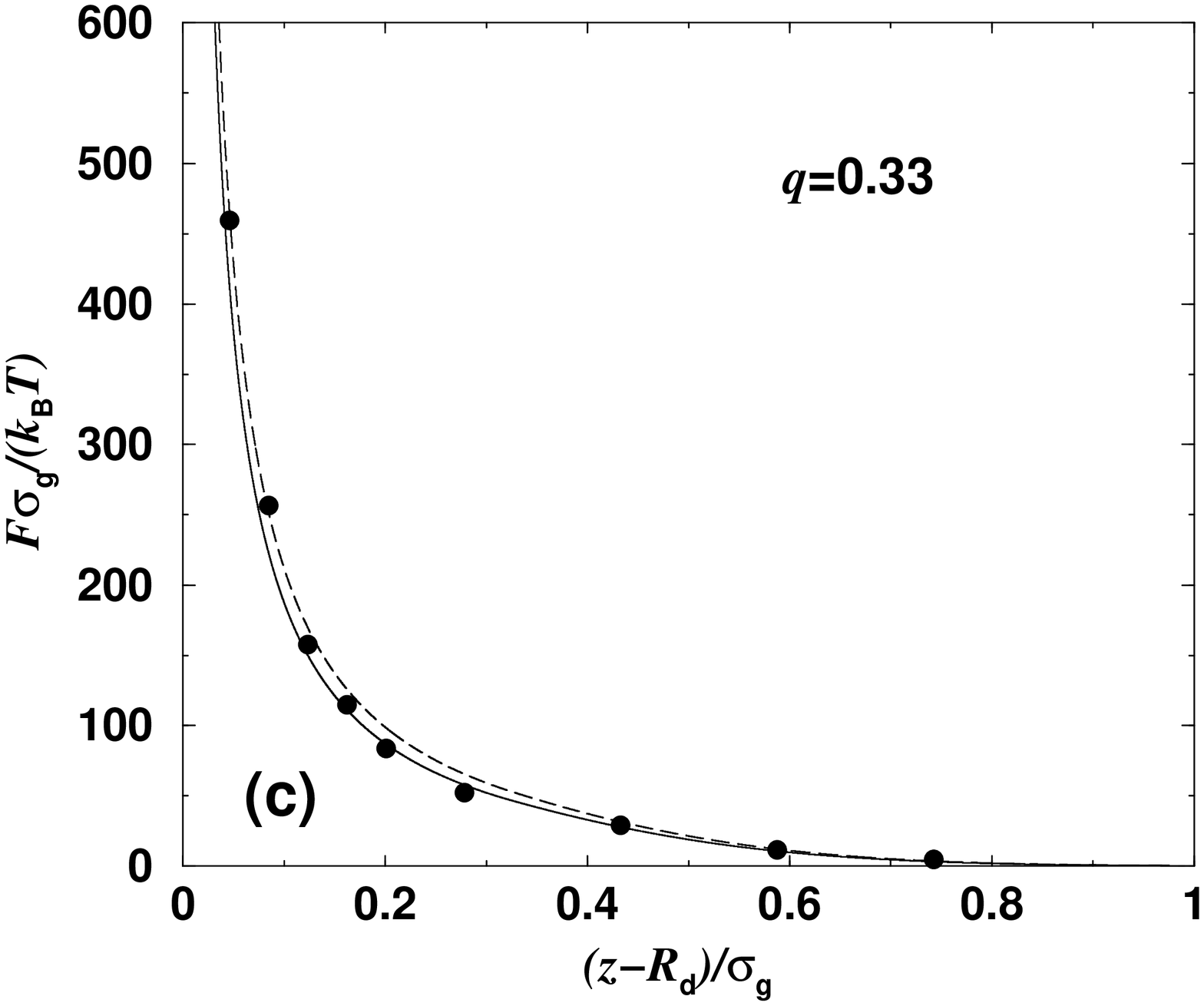}
\includegraphics[width=10.0cm,angle=-90]{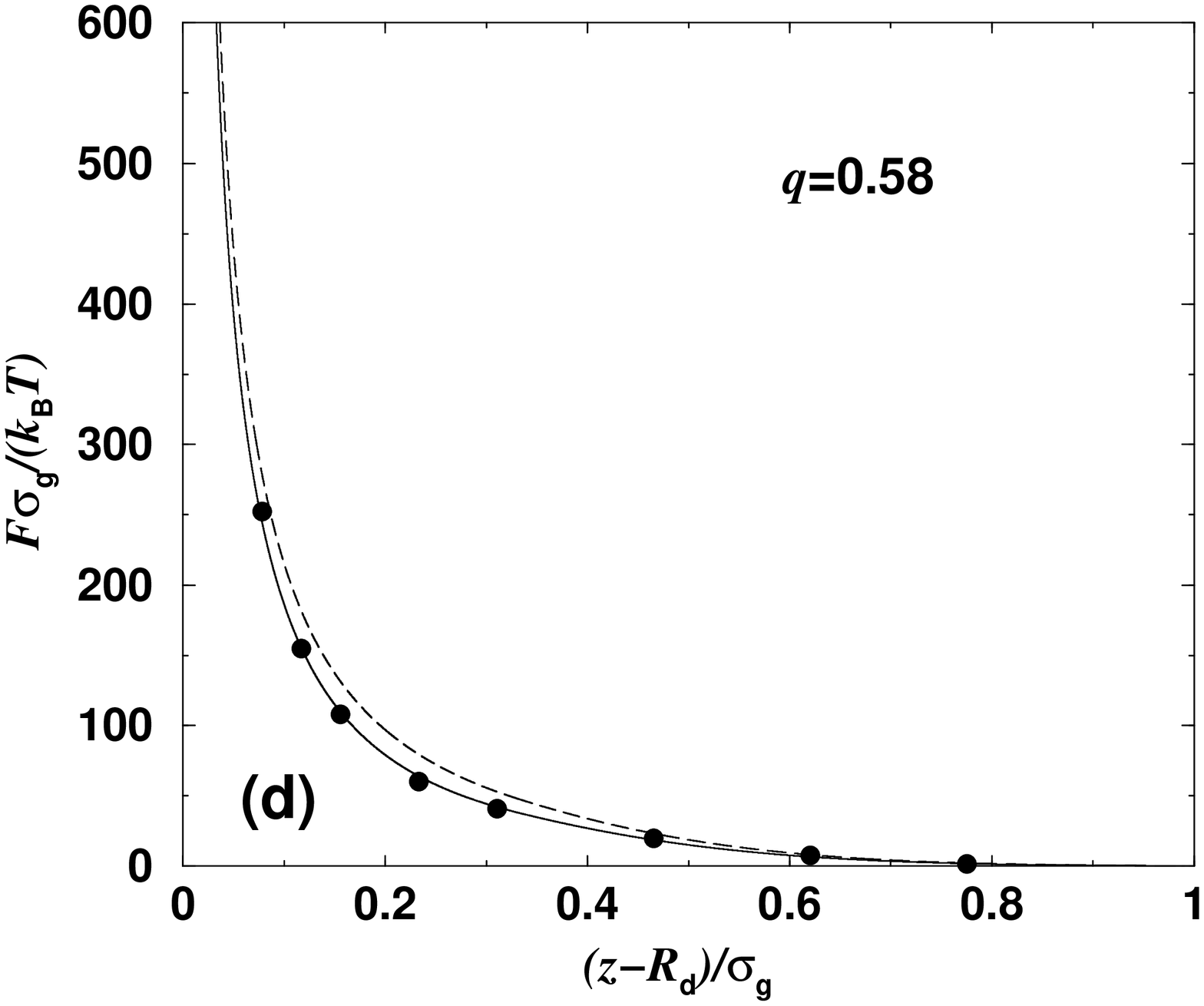}
\end{center}
\end{figure}
\clearpage

\begin{figure}
\begin{center}
\includegraphics[width=10.0cm,angle=-90]{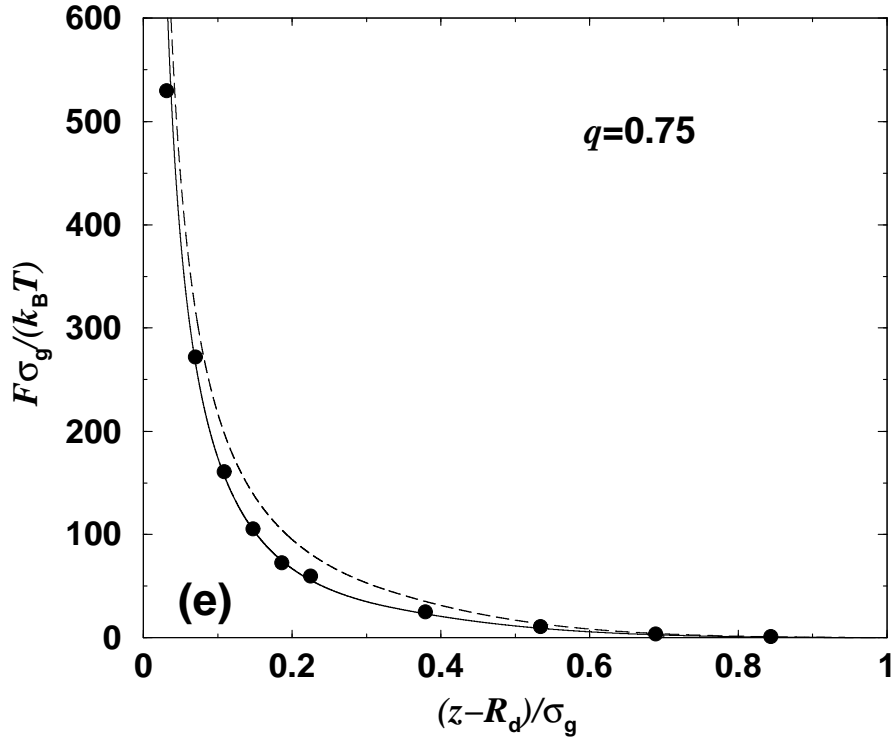}
\end{center}
\caption{Comparison between simulation (symbols) and theoretical (lines) 
  results for the effective force between a star
  polymer and a colloidal particle for different size ratios $q$,
  as a function of the center-to-surface separation $z$.
  The arm number here is $f = 18$.
  The solid lines in (a) and (b) are
  derived from Eq.\ (\ref{fsc.eq}) for $s_{\rm max} \rightarrow \infty$.
  In (c)-(e) the curves derived by means of this approximation are
  shown dashed and they increasingly deviate from the simulation results
  as $q$ grows. Thereby, a finite upper integration limit has to
  be introduced (see the text), producing the curves denoted
  by the solid lines in (c)-(e) and bringing about excellent agreement
  with simulation.}
\label{star-coll_18}
\end{figure}

\begin{figure}
\begin{center}
\includegraphics[width=15.0cm,angle=-90]{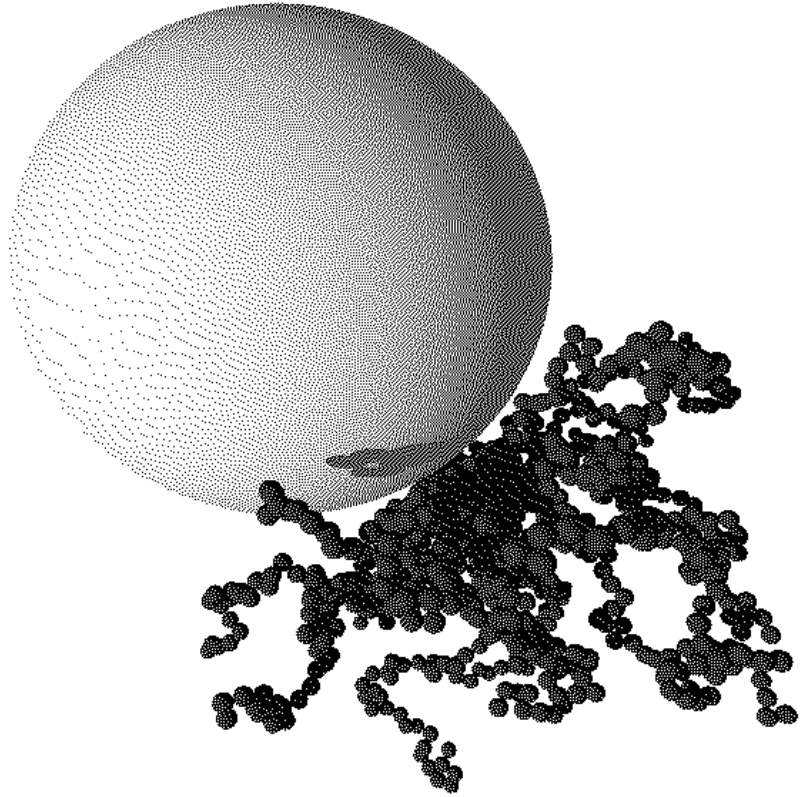}
\end{center}
\caption{Snapshot of a typical configuration of a star polymer 
         with $f=18$ arms near a colloidal sphere with $q=0.75$.
         One should notice that predominantly the
         inner region of the star interacts with the hard
         sphere, yielding the main contribution of the inner core regime to
         the osmotic pressure of a region, determined by 
         $\theta_{\rm max}\approx 30^{\rm o}$. Thereby, the upper
         integration limit $s_{\rm max}$ in Eq.\ (\ref{Fz.eq}) is limited,
         see also the geometrical aspects of Eq.\ (\ref{rmax.eq}) 
         and Fig.\ \ref{star-colloid}.}
\label{pov}
\end{figure}

\begin{figure}
\begin{center}
\includegraphics[width=12.0cm,angle=-90]{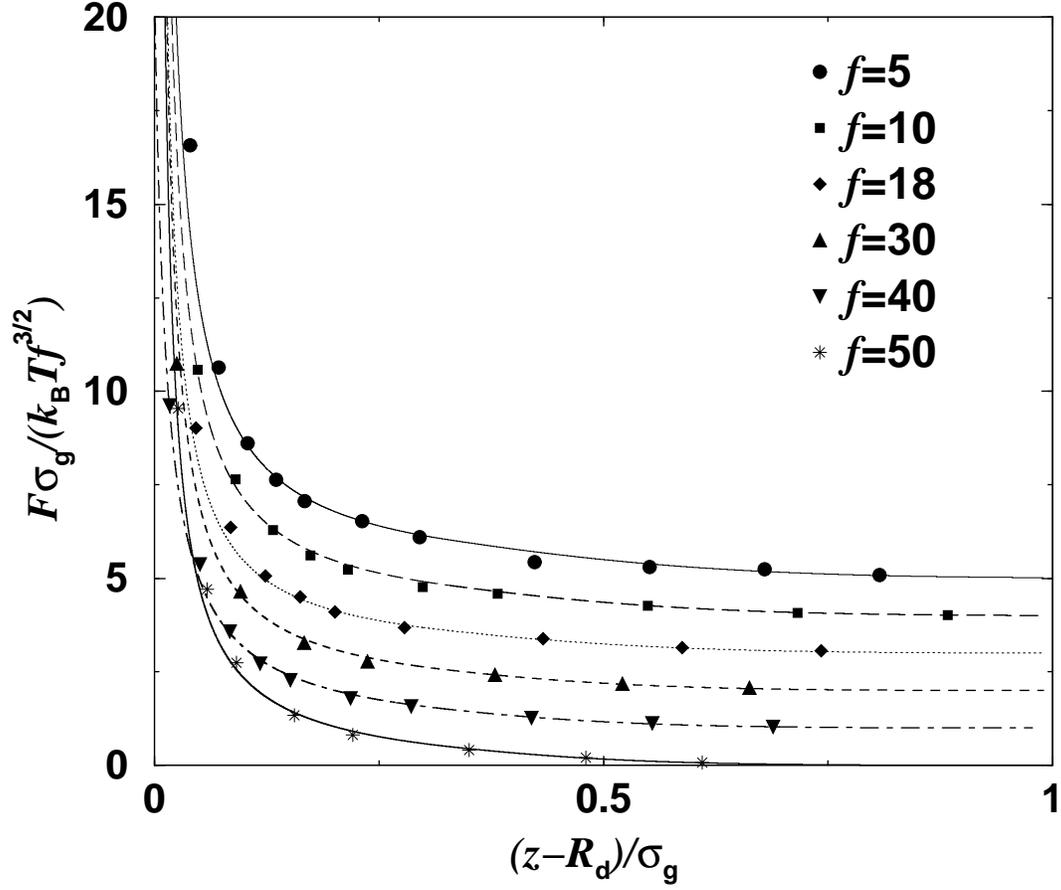}
\end{center}
\caption{The effective force between a star polymer and a colloid for
 different arm numbers $f$ and $q =  0.33$ plotted against $z$, the distance
 of the star center to the surface of the colloid. 
 The lines are the theoretical and the symbols the simulation results.
 For clarity,
         the data have been shifted upwards by constants: $f = 10:1$,
         $f = 18:2$, $f = 30:3$, $f = 40:4$, $f = 50:5$.}
\label{star-coll_0.3}
\end{figure}

\begin{figure}
\begin{center}
\includegraphics[width=12.0cm,angle=-90]{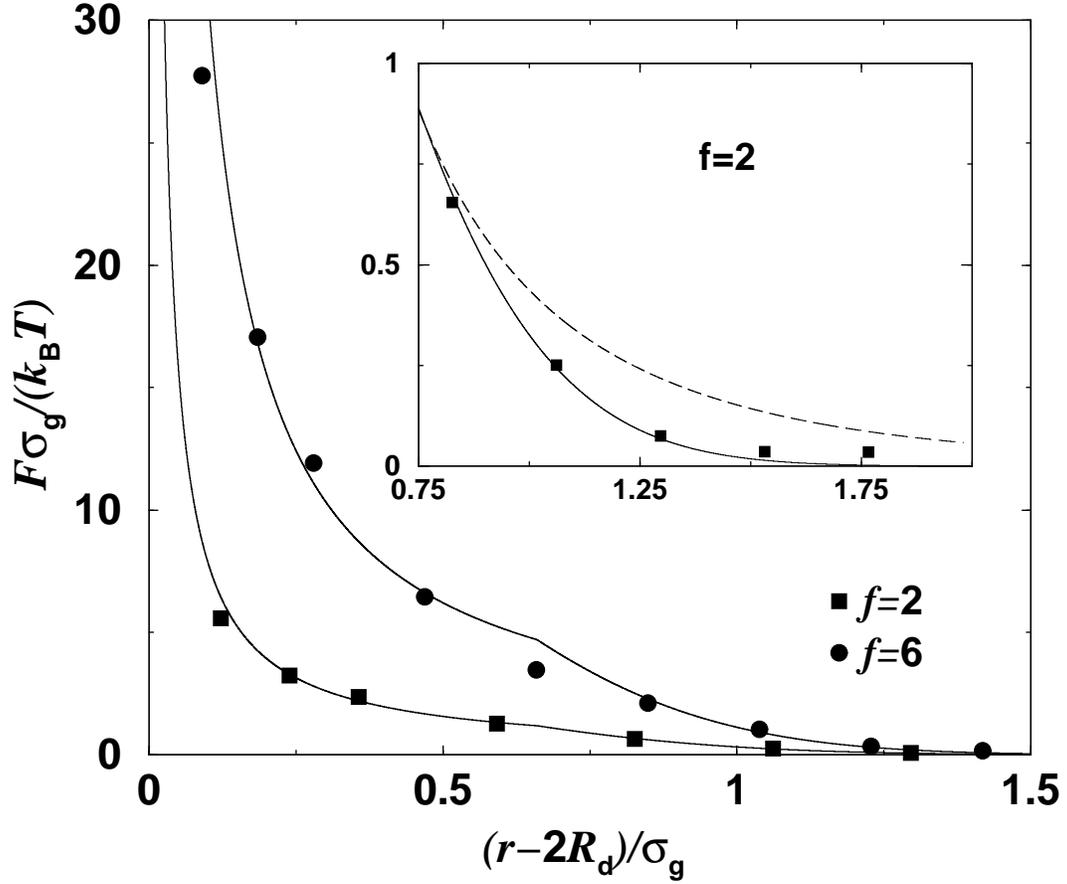}
\end{center}
\caption{Effective force between two star polymers plotted against the
  center-to-center distance $r$ for arm numbers $f=2$ and $f=6$. The
  simulation data (symbols) coincide with the logarithmic-Gauss expression
  (solid lines) of Eq.\ (\ref{pot_ss2}). 
  In the inset, the outer
  distance region is enlarged in order to clearly show the
  validity of the Gaussian decay in this $f$-regime (solid line),
  whereas the Yukawa
  form (dashed line) produces poor agreement there.}
\label{2stars-inset}
\end{figure}

\begin{figure}
\begin{center}
\includegraphics[width=12.0cm,angle=-90]{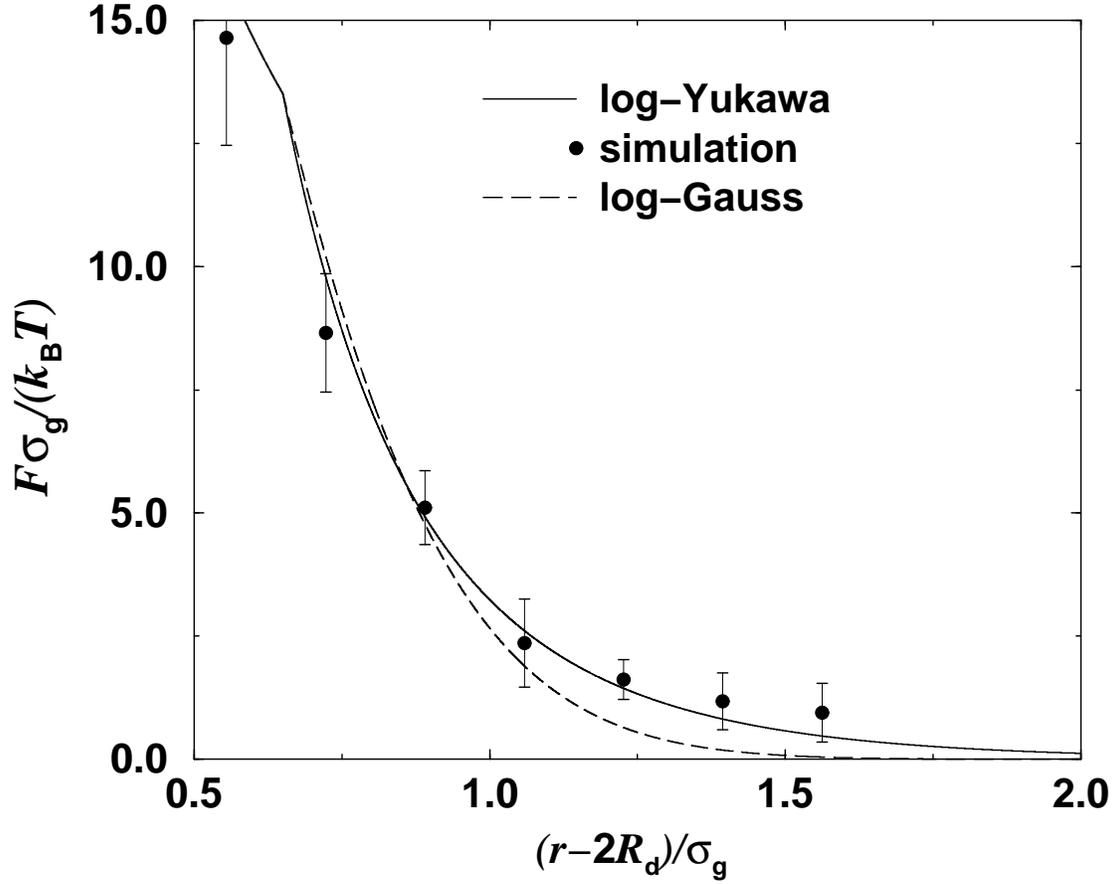}
\end{center}
\caption{Effective force between two star polymers plotted against the
  center-to-center distance $r$ for arm number $f = 10$. In contrast
  to Fig.\ \ref{2stars-inset}, here the Yukawa form (solid line) 
  gives an accurate
  description of the decay of the interaction at large separations,
  whereas the Gaussian form (dashed line) does not.}
\label{2stars-f10}
\end{figure}

%\clearpage
%
\end{document}